\documentclass[%
 reprint,
 amsmath,amssymb,
 aps,
]{revtex4-2}

\usepackage{blindtext}
\usepackage{graphicx}
\graphicspath{{Figures/}}
\usepackage{siunitx}
\usepackage{chemformula}
\usepackage{hyperref}

\begin{document}

\preprint{APS/123-QED}

\title{Plasma cleaning of LCLS-II-HE verification cryomodule cavities}

\author{B. Giaccone}
\email{giaccone@fnal.gov}
\author{P. Berrutti}
\author{M. Martinello}
\author{S. Posen}
\author{A. Cravatta}
\author{A. Netepenko}
\author{T. Arkan}
\author{A. Grassellino}
\author{B. Hartsell}
\author{J. Kaluzny}
\author{A. Penhollow}
\affiliation{%
Fermi National Accelerator Laboratory, Kirk Rd and Pine St, Batavia, Illinois 60510, USA
}%

\author{D. Gonnella}
\author{M. Ross}
\author{J. Fuerst}
\author{G. Hays}
\author{J. T. Maniscalco}

\affiliation{%
SLAC National Accelerator Laboratory, 2575 Sand Hill Rd, Menlo Park, California 94025, USA
}%

\author{M. Doleans}
\affiliation{%
Oak Ridge National Laboratory, 1 Bethel Valley Rd, Oak Ridge, TN 37830 USA
}%

\date{\today}

\begin{abstract}
Plasma cleaning is a technique that can be applied in superconducting radio-frequency cavities \textit{in situ} in cryomodules to decrease their level of field emission. We developed the technique for Linac Coherent Light Source II (LCLS-II) cavities and we present in this paper the full development and application of plasma processing to the LCLS-II High Energy verification cryomodule (vCM).
We validated our plasma processing procedure on the vCM, fully processing four out of eight cavities of this CM, demonstrating that cavity performance was preserved in terms of both accelerating field and quality factor. Applying plasma processing to this clean, record breaking cryomodule also showed that no contaminants were introduced in the string, maintaining the vCM field emission-free up to the maximum field reached by each cavity. We also found that plasma processing eliminates multipacting (MP)-induced quenches that are frequently observed within the MP band field range. This suggests that plasma processing could be employed \textit{in situ} in CMs to mitigate both field emission and multipacting, significantly decreasing the testing time of cryomodules, the linac commissioning time and cost and increasing the accelerator reliability.

\end{abstract}

\maketitle


\section{\label{sec:intro}Introduction}
The term plasma cleaning refers to a process in which impurities are removed from a surface via a mixture of inert and reactive gases in the form of a plasma. This technique, also called plasma processing, was developed for superconducting radio-frequency (SRF) cavities some years ago by Oak Ridge National Laboratory (ORNL). The developers demonstrated that a plasma composed of a mixture of neon and a small percentage of oxygen reduced hydrocarbon-related field emission in the Spallation Neutron Source (SNS) cavities \cite{doleans2013plasma, doleans2016ignition, doleans2016plasma, doleans2016situ}. Starting from this successful experience, plasma processing studies are being conducted at multiple laboratories for different accelerating structures \cite{giaccone2021splasma, berrutti2019plasma, giaccone2021field, powersplasma, zhangplasma, wu2018situ, wu2019cryostat, huang2019effect}.

Plasma processing for \SI{1.3}{\giga\hertz} SRF cavities was developed at Fermi National Accelerator Laboratory (FNAL) \cite{berrutti2018, berrutti2019plasma, giaccone2021field}, in collaboration with SLAC National Accelerator Laboratory and ORNL with the purpose of  mitigating field emission \cite{fowler1928electron, padamseerf} in the \SI{1.3}{\giga\hertz} Linac Coherent Light Source II (LCLS-II) \cite{lclsIIdesign} and LCLS-II High Energy (LCLS-II-HE) \cite{schoenlein2016lcls} cavities.

\begin{figure}
    \centering
    {\includegraphics[width=1\columnwidth]{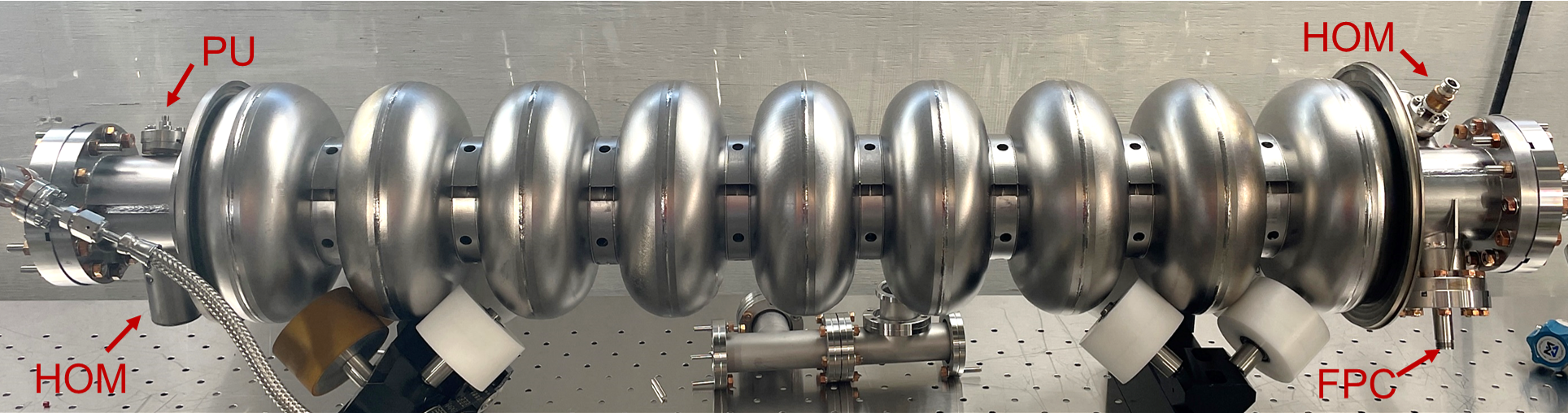}} \quad
	\caption{Photo of TESLA shape 9-cell \SI{1.3}{\giga\hertz} cavity, with highlighted ports for the higher order modes (HOM) couplers, the fundamental power coupler (FPC) and the pick up (PU) coupler.}
	\label{fig:TESLAcavity}
\end{figure}

As discussed in Berrutti \textit{et al.} \cite{berrutti2019plasma}, a new method of plasma ignition was developed for LCLS-II \SI{1.3}{\giga\hertz} cavities. This innovative method utilizes the higher order modes (HOM) and HOM couplers (Fig. \ref{fig:TESLAcavity}) to ignite and move the glow discharge inside the cavity RF volume. A systematic study of plasma processing applied to LCLS-II \SI{1.3}{\giga\hertz} nitrogen-doped \cite{grassellino2013nitrogen} cavities was carried out at FNAL, demonstrating that plasma cleaning can successfully mitigate hydrocarbon-related FE without affecting the high Q-factors and quench fields characteristic of N-doped cavities \cite{giaccone2021field}.

In this paper the development and first application of plasma processing to a full size \SI{1.3}{\giga\hertz} cryomodule (CM)  are presented. 

The cryomodule under study is the LCLS-II-HE verification cryomodule (vCM). Before plasma processing the vCM underwent extensive tests, and the results are reported in Posen \textit{et al.} \cite{posen2021lcls}. The LCLS-II-HE project requires \SI{1.3}{\giga\hertz} 9-cell niobium cavities operating at quality factor $\mathrm{Q_0} > 2.7 \times 10^{10}$ and accelerating gradient $\mathrm{E_{acc}} \approx 21\,$MV/m. To reach these specifications, a new N-doping treatment was developed, called `2/0' since cavities are treated with nitrogen for only two minutes and subsequently cooled down in the furnace \cite{gonnella2019lcls}. The vCM showed record performance in terms of both quality factor and accelerating gradient, exceeding the project specification. Particularly relevant for plasma processing, only one cavity (CAV5) exhibited detectable radiation ($0.6\,$mR/hr), but the FE source was processed during later testing, leaving the vCM field emission-free. Nevertheless, the vCM represented a unique opportunity to scale the \SI{1.3}{\giga\hertz} plasma processing technique from a single TESLA-style \cite{aune2000superconducting} cavity to an entire cryomodule. For this reason, we plasma processed four out of eight cavities of the vCM and we compared the RF cavity performance before and after plasma processing. The details of the development of the technique from the single cavity system to the cryomodule are presented, together with the risk analysis and mitigation strategy that was conducted. We also summarize the plasma processing plan, the details of the experimental system, and the RF test results. Of particular interest is the multipacting (MP) quench analysis that was conducted before and after plasma processing. The four cavities that underwent plasma cleaning did not show any MP quenches, demonstrating that this treatment can fully eliminate multipacting in cavities in cryomodules. Therefore, performing plasma processing after a cryomodule has been assembled may significantly decrease its overall testing time and the subsequent accelerator commissioning time, thus reducing the overall project cost. In addition, cryomodules with multipacting-free cavities would offer greater stability during the accelerator operation, increasing its reliability.

\section{\label{sec:risk} Risk Analysis and Mitigation}

All major risks were identified before deploying plasma processing in a fully assembled and tested cryomodule. The main risks that were identified are the following: i) potential damage of the HOM cables, connectors and feedthroughs in the module due to the RF power passing through them during plasma ignition; ii) the potential ignition of plasma in the fundamental power coupler (FPC) and consequent damage of the FPC; and iii) potential pressure instabilities throughout the cryomodule.

\begin{figure}
	\centering
	{\includegraphics[width=0.9\columnwidth]{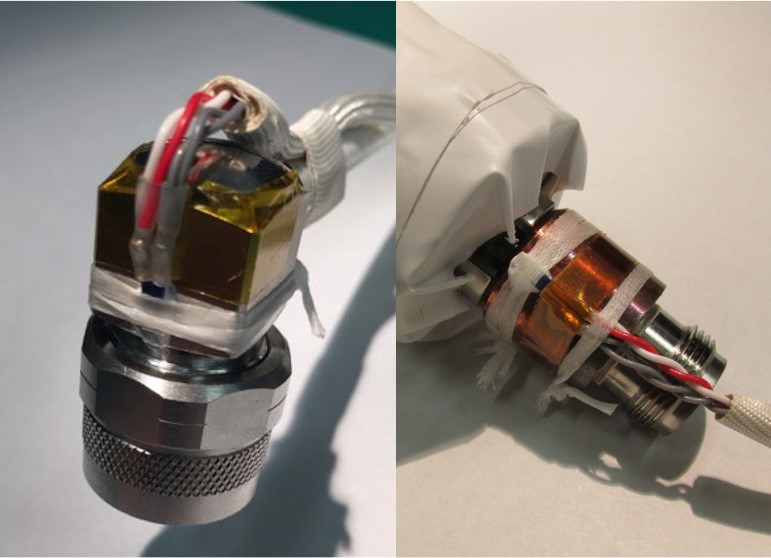}} \quad
	\caption{Temperature sensors installed on the cable end (left) and cavity HOM feedthroughs (right).}
	\label{fig:temp_sens_cf}
\end{figure}

\begin{figure}
	\centering
	{\includegraphics[width=1\columnwidth]{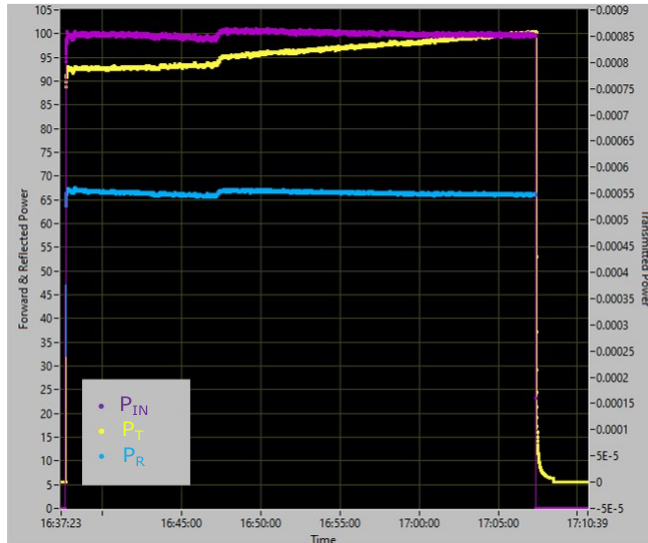}} \quad
	\caption{Power versus time during cable testing for LCLS-II plasma processing: input power of \SI{100}{\watt} for 30 minutes. The input and reflected power (respectively $\mathrm{P_{IN}}$, $\mathrm{P_{R}}$) follow the y-axis, while the right y-axis refers to the transmitted power $\mathrm{P_{T}}$.}
	\label{fig:PW_cable_test}
\end{figure}

\begin{figure}
	\centering
	{\includegraphics[width=0.95\columnwidth]{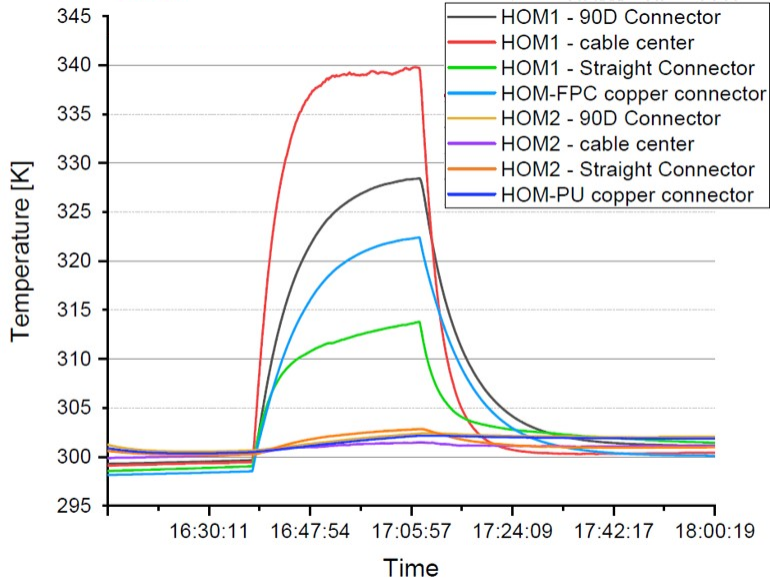}} \quad
	\caption{Temperature of HOM1 and HOM2 cables and feedthrus versus time during cable testing for plasma processing: input power of \SI{100}{\watt} for 30 minutes.}
	\label{fig:Temp_cable_test}
\end{figure}

To verify the first point, HOM cables were tested extensively by subjecting them to different levels of input power, from \SI{10}{\watt} to \SI{125}{\watt}.
The heat and temperature distributions along the input and transmitted power cables were monitored through temperature sensors installed on the cable connectors and at the cable center point. Another set of temperature sensors was installed on the cavity HOM feedthroughs in which there are fragile components, such as the ceramic window. An image of the temperature sensors installed on a cable end and HOM feedthrough is shown in Fig. \ref{fig:temp_sens_cf}.
The RF characteristics of the cables were measured before and after each test, and both the reflected ($S_{11}$) and transmitted ($S_{21}$) signals were recorded. The measurements of both $S_{11}$ and $S_{21}$ did not change after any of the tests, therefore, the cable performance was not affected by the RF power. The frequency chosen for these tests was \SI{1829}{\mega\hertz}. This frequency was chosen to be close to the mode used to ignite the plasma in LCLS-II cavities and, to maximize the amount of reflected power within a MHz range from the cavity RF mode. This ensured exceeding the reflected power and the heat dissipated in the cables under regular plasma conditions: during cavity processing, power is sent to the resonator at its mode frequency, thus minimizing reflected power.
The most significant cable test was conducted at \SI{100}{\watt} of input power for 30 minutes: \SI{100}{\watt} indeed corresponds to the power required for plasma ignition. That level of power is needed only for a few seconds at ignition, and $10\,-\,$\SI{20}{\watt} of RF power is sufficient to sustain the plasma afterwards. Details regarding input, reflected and transmitted power during plasma cleaning are discussed in section \ref{sec:PlasmaData}.
The input power versus time is shown in Fig. \ref{fig:PW_cable_test} and the temperature profile is presented in Fig. \ref{fig:Temp_cable_test}. The maximum cable temperature increase, of approximately \SI{44}{\kelvin}, does not damage or change the RF performance of the cables. One can observe how the central point of the input cable is the hottest spot, followed by the end connector, the cavity HOM feedthrough and lastly the cable connector to the power amplifier. As expected, the difference in cable temperature between the input (HOM1) and the transmitted (HOM2) power sides is quite large: the HOM2 cable temperature increase is well below \SI{5}{\kelvin}. 

To verify whether plasma ignition in the FPC represented a risk during plasma processing in a cryomodule, an experiment was carried out on a 9-cell cavity on a test bench: a variable coupler was installed on the cavity, with identical antenna tip design and coupling to the cavity as the production FPC for cryomodule operation. The cavity was then plasma processed following the standard procedure. The results of this test were encouraging because plasma ignition in the coupler volume was not observed. In addition, after the test, both the input HOM antenna and the FPC were optically inspected and found to be free of any kind of damage or discoloration.

Last, to minimize the pressure fluctuations along the beamline vacuum portion of the cryomodule, a new vacuum cart was assembled and tested on a single cavity prior to plasma processing the vCM.

\section{\label{sec:TestPlan} Experimental setup and plasma processing procedure}

Plasma cleaning of the vCM took place at the Cryomodule Test Facility (CMTF) at FNAL after the completion of extensive RF tests, the results of which are discussed in Posen \textit{et al.} \cite{posen2021lcls} and the subsequent warm up to room temperature. The vCM was RF tested again after plasma cleaning to assess the cavity performance changes after plasma processing.

\begin{figure}
	\centering
	{\includegraphics[width=1\columnwidth]{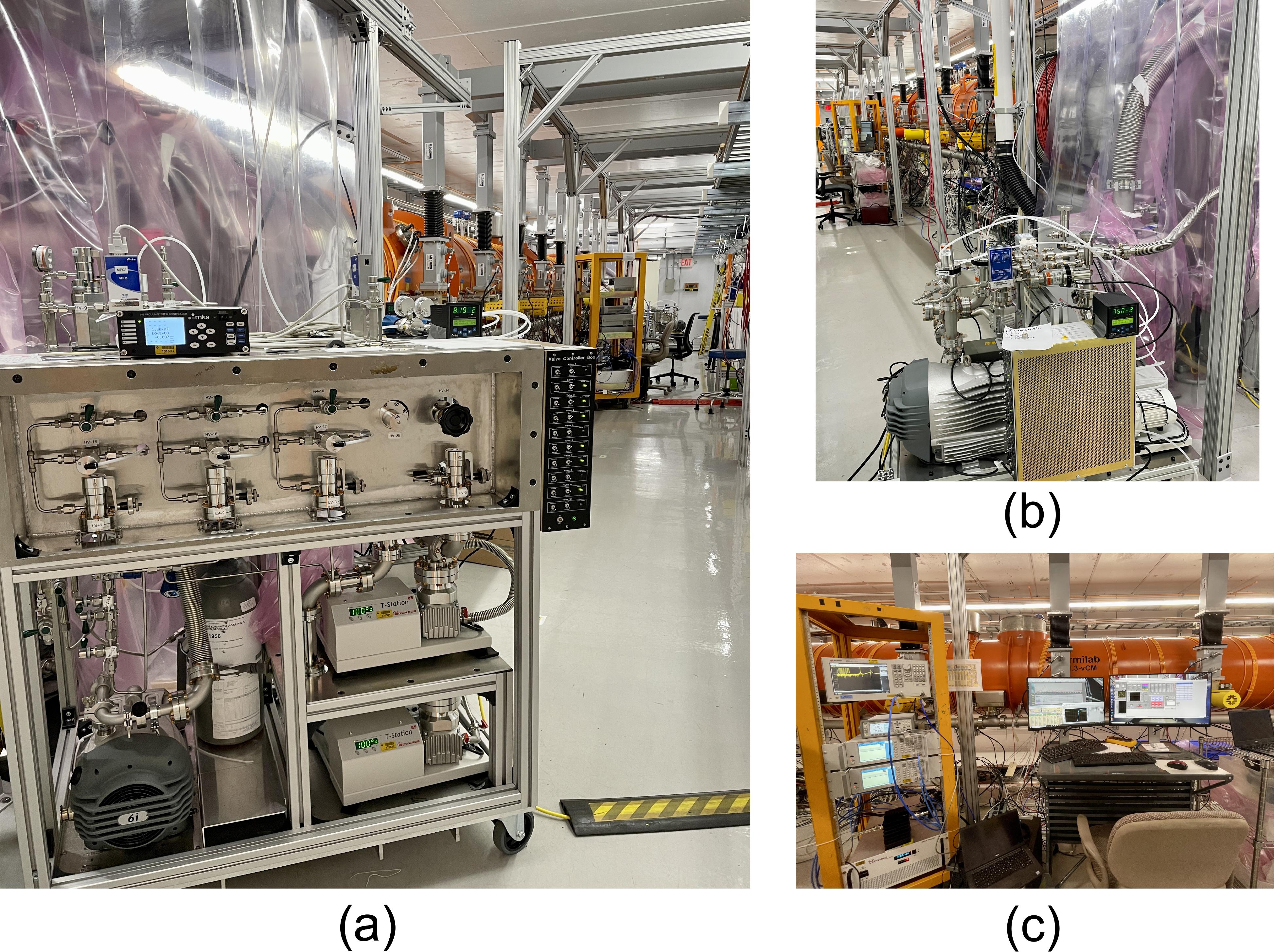}} \quad
	\caption{Experimental setup used to apply plasma processing to the LCLS-II-HE vCM. The gas station is connected on one end of the CM (Fig. (a)), the vacuum and RGA station to the other end (Fig. (b)). The RF rack and computers are located in between the gas and vacuum stations (Fig. (c)).}
	\label{fig:experimentalsetup}
\end{figure}

Photographs of the experimental system used to plasma process the vCM are shown in Fig. \ref{fig:experimentalsetup}. A gas injection system was connected to the upstream side of the vCM (near CAV1), and a vacuum cart was connected to the downstream side (near CAV8). The connections between the gas and vacuum carts and the vCM were carried out in a class-100 portable cleanroom to minimize the risk of particle contamination. Both gas and vacuum carts are equipped with pneumatic valves that guarantee instantaneous and easy operation of the two carts, in addition to allowing for future automation of the systems. All-range pressure gauges were used to monitor the pressure of the gas at the two ends of the CM. The gas mixture and plasma by-products partial pressures are measured with a residual gas analyzer (RGA) located in the vacuum cart, downstream of the cavity string. The percentage of oxygen contained in the gas mixture can be manually adjusted based on the partial pressures measured by the RGA and is usually maintained at approximately $1-2\%$ of the neon value. A complete description of the RF system and procedure used for plasma ignition and plasma moving across the cavity can be found in \cite{berrutti2019plasma, giaccone2021field}. The power injected and reflected on the HOM1 (HOM used as input antenna) and transmitted by the HOM2 (HOM used as field probe antenna) is constantly monitored and recorded during plasma processing. A LabVIEW program was used to monitor and record the gas pressure and the power levels and to operate the two RF signal generators to ignite and transfer the plasma from cell to cell. In addition, the software was also connected to the network analyzer (VNA), making it possible to automatically identify the cavity resonance frequencies of the two dipole pass-bands used for plasma processing and to detect in which cell plasma ignition has occurred.

Plasma cleaning was carried out in four out of the eight cavities of the vCM. This enabled the ability to i) compare the performance of cavities subjected (and not subjected) to plasma cleaning, and to ii) perform cleaning within the time constraint imposed by the project. To monitor the cavity's surface temperature during processing, the four cavities with additional temperature sensors installed inside the helium vessel were chosen for plasma processing, namely, CAV1, CAV4, CAV5, and CAV8 (the number indicates the position along the cavity string). These four cavities were instrumented with four temperature sensors placed on the outside wall of the cavity on cells $\#$ 1 and $\#$ 9 at the top and bottom positions. The same cavities were also instrumented with temperature sensors on the HOM connectors, offering the advantage of being able to monitor the temperature of both higher order mode cables. In addition, CAV1 and CAV5 each have two more temperature sensors placed on the outside of the helium vessel.

During plasma cleaning, a mixture of neon and oxygen in gaseous form is injected into the cavity string, and a constant flow is maintained from CAV1 to CAV8. The glow discharge is ignited in one cell at a time following the direction of the gas flow; the cavities were also processed following the direction of the gas flow from CAV1 to CAV8. As explained in Giaccone \textit{et al.} \cite{giaccone2021splasma}, two identical passes of plasma cleaning are applied to each cavity. Due to uncertainty regarding the time available to apply plasma processing to the vCM, CAV1 subsequently received the two runs of plasma processing, meaning that as soon as the first pass was completed, the second started. CAV4, 5 and 8 instead received the first plasma pass one after the other, then the second run restarted from CAV4 to 8. 

In addition to the cavity and HOM cable temperature, the pressure of the gas mixture, the partial pressure of oxygen and the plasma byproducts (\ch{C}, \ch{CO}, \ch{CO2}, \ch{H_2O}), the power levels (input, reflected and transmitted) and the cavity spectrum were monitored.

\section{\label{sec:PlasmaData} Data analysis and discussion}

The vCM cavity string was maintained under high vacuum during and after the connections to the plasma processing gas and vacuum carts. Before starting the cavity plasma cleaning, the gas flow was established through the cavity string and the pressure was slowly increased up to the working point ($\approx 134\,$mTorr). Once the pressure stabilized, the pressure drop across the two ends of the vCM was measured to be $\approx 24\,$mTorr. The relative concentrations of neon and oxygen in the gas mixture were monitored with the RGA and adjusted as necessary by operating the gas cart leak valves controlling the mixture of the pure neon and the $20\%$ oxygen balance neon cylinders. Once the \ch{O_2} partial pressure stabilized to the desired level (between $1.5-2\%$), the glow discharge was ignited inside the cavity RF volume.

\begin{figure}[h!]
	\centering
	{\includegraphics[width=1\columnwidth]{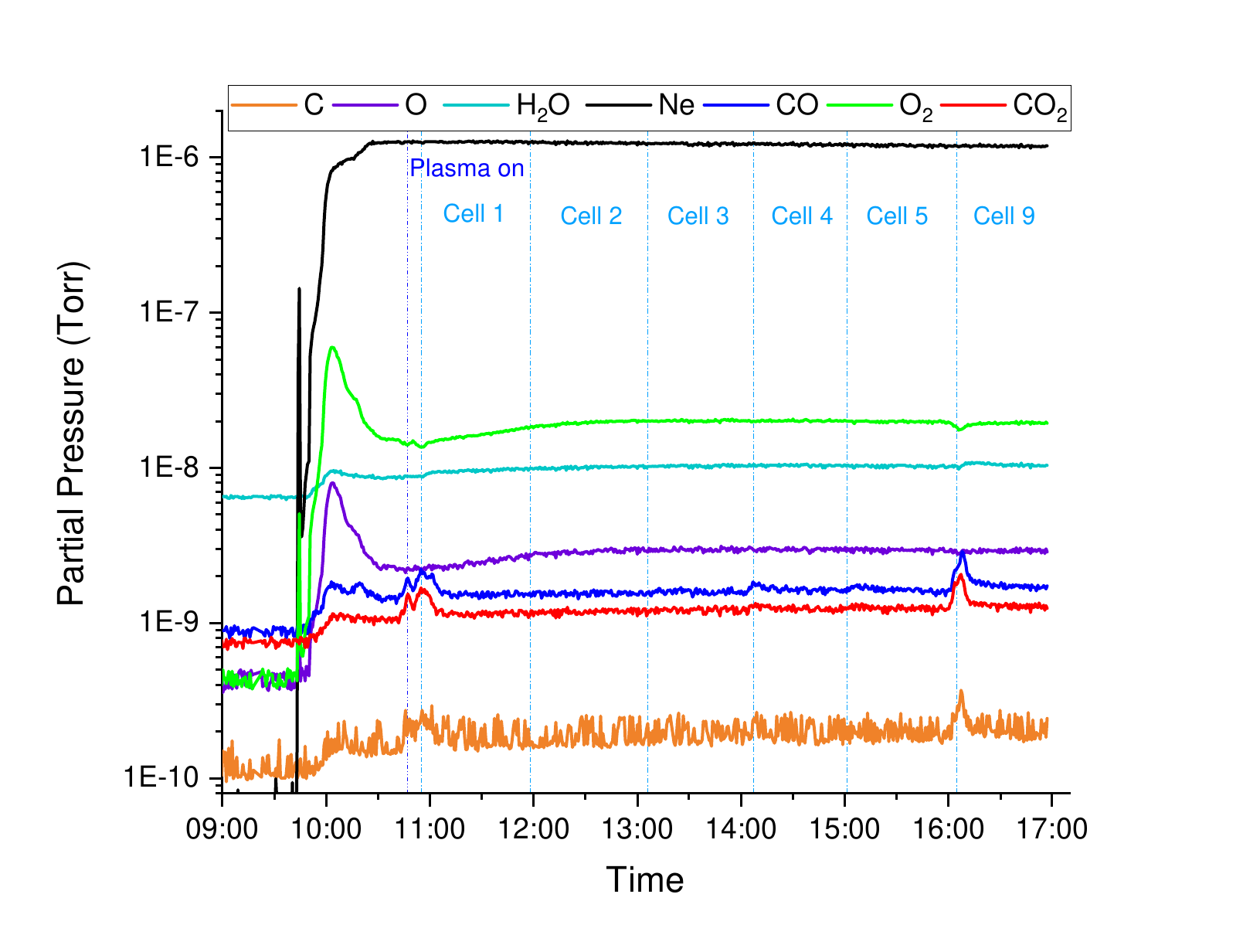}} \quad
	\caption{Plot of RGA signals measured on CAV1, during the first day of plasma processing from cell $\#$ 1 to $\#$ 5 and cell $\#$ 9. As the gas flow is started and the pressure in the vCM increases, there is an initial peak in the oxygen signals (at 16 and 32 amu), which then decrease and stabilize to $\approx 1.6 \%$ of the neon content.}
	\label{fig:RGAData_cav1}
\end{figure}

The RGA data collected during the first run of plasma processing on CAV1 are shown in Fig. \ref{fig:RGAData_cav1}. Once the glow discharge ignites inside the cavity, the carbon-related signals (\ch{CO}, \ch{CO_2}, \ch{C}) increase as shown in Fig. \ref{fig:RGAData_cav1}. The subsequent peak corresponds to the transfer of the plasma and the tuning of its density in cell $\#$ 1. Concurrently, a small decrease in the \ch{O_2} level can also be observed, confirming that the reactive oxygen ions present in the glow discharge are creating volatile byproducts with the hydrocarbons knocked off the cavity surface. However, as shown in Fig. \ref{fig:RGAData_cav1}, not all cells showed a significant change in the RGA signals after the plasma was transferred. Only in this first plasma cleaning run was the processing conducted from cell $\#$ 1 to 9, while in all subsequent cleaning runs it was conducted from cell $\#$ 9 to 1, following the direction of the gas flow. 

At the end of each plasma cleaning session (usually corresponding to the end of the day), the gas flow was stopped, and the cavity string was evacuated using the plasma vacuum cart. Then, the cavity string was isolated by manually closing the two right angle valves connecting the string to the gas and vacuum carts. During these downtimes, the pressure in the cavity string was monitored using beamline pressure gauges.

\begin{figure}
	\centering
	{\includegraphics[width=1\columnwidth]{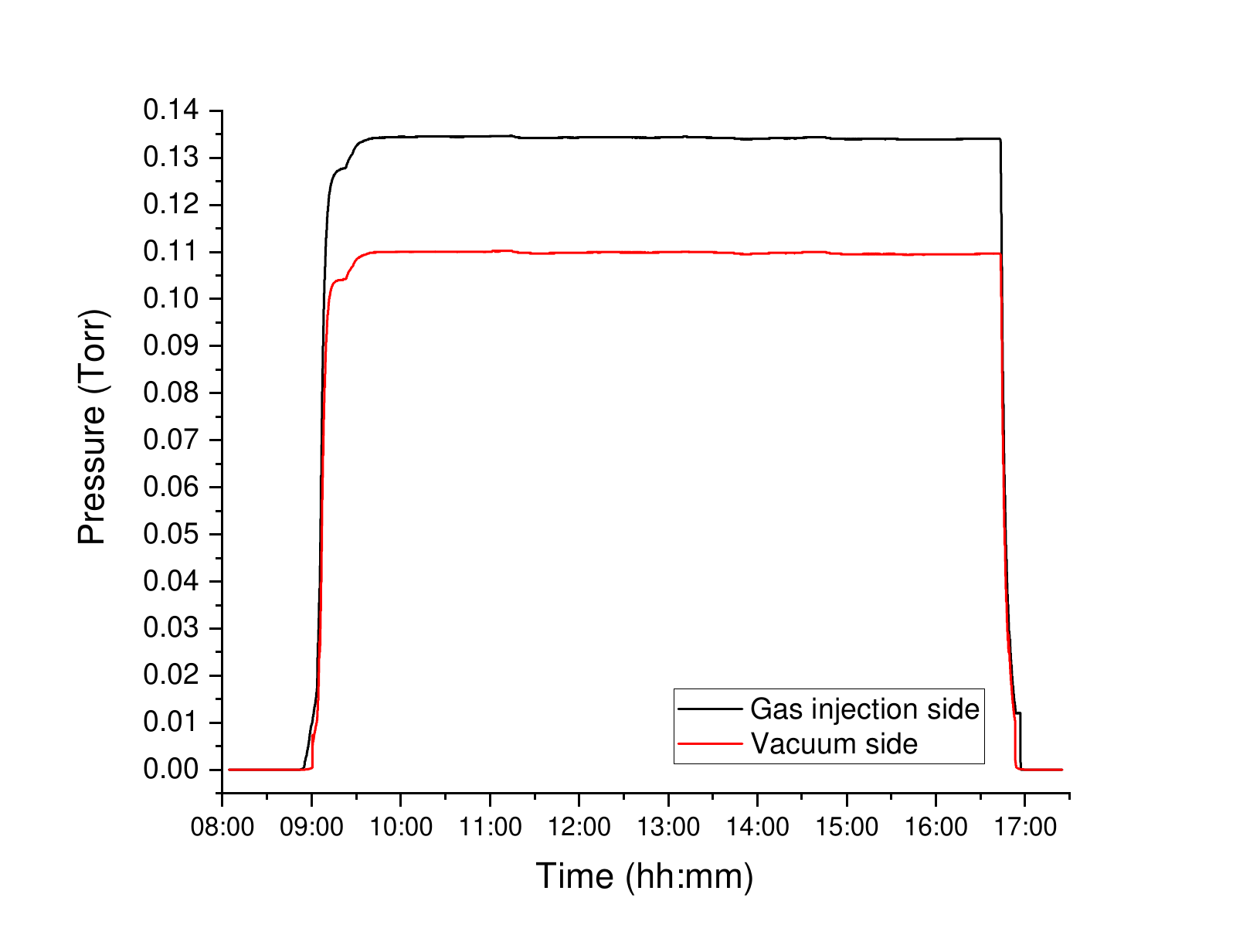}} \quad
	\caption{Pressure measured at the two ends of the vCM during initial neon-oxygen mixture injection and during the flow maintained for the entire duration of plasma processing. The pressure drop across the string is approximately $24\,$mTorr. The pressure values are scaled to account for the gauge sensitivity to neon. These data were acquired during plasma processing of CAV4.}
	\label{fig:GasPressure_cav4}
\end{figure}

\begin{figure*}
	\centering
	{\includegraphics[width=1\textwidth]{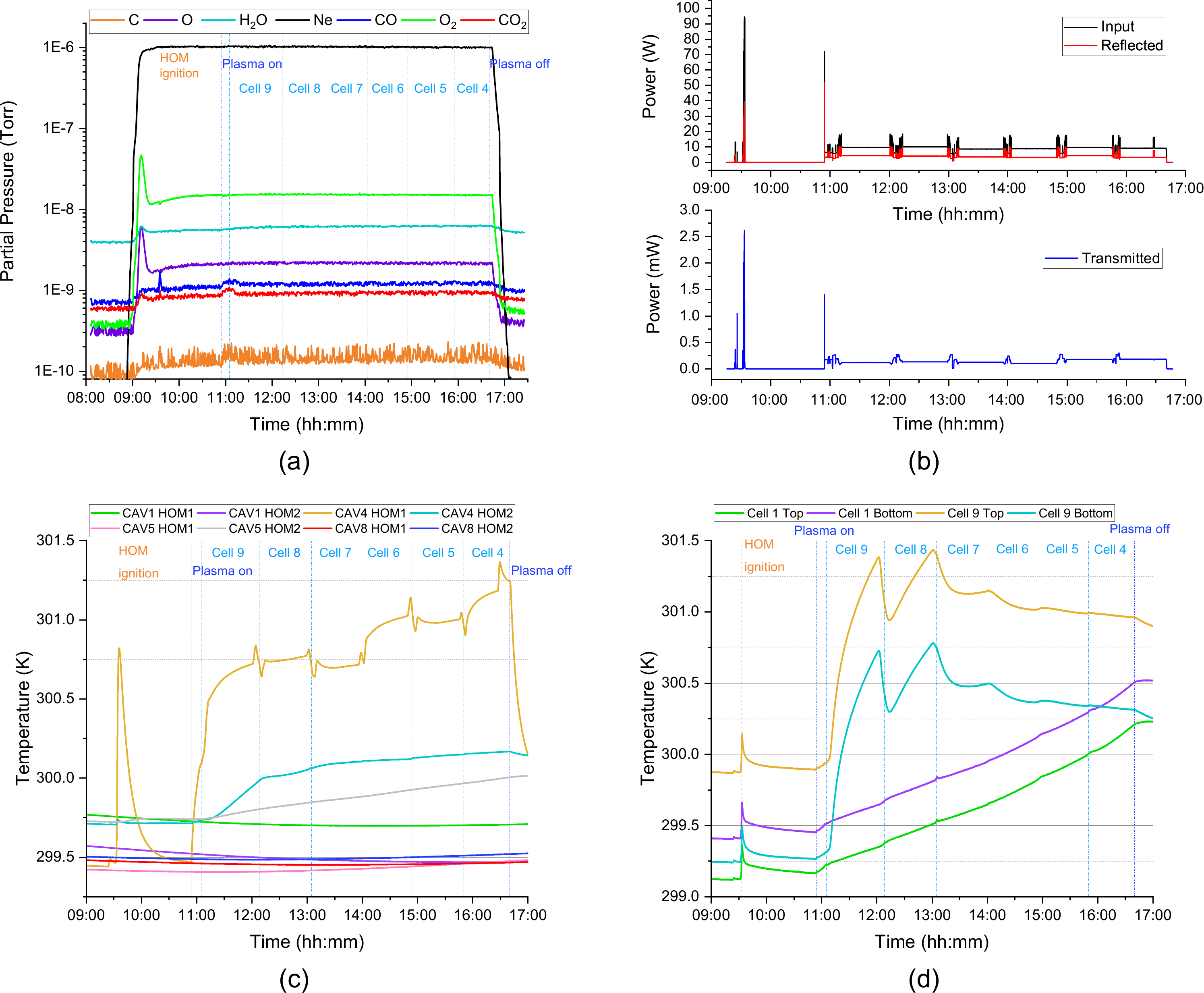}} \quad
	\caption{Experimental data collected during plasma processing of CAV4. Panel (a) contains a plot of the RGA signals measured during plasma processing of cells $\#$ 9 to $\#$ 4. The peak at approximately 9:30 measured for \ch{CO} and \ch{CO_2} indicates the accidental ignition of the plasma in the HOM coupler. The following peak at approximately 11:00 is measured when the plasma is successfully ignited inside the cavity. Panel (b) shows the input and reflected power measured at the CAV4 HOM1 coupler and transmitted by the HOM2 during plasma processing. Note the different scales on the y-axis between the top and bottom plots. Additionally in this case, the 9:30 peak corresponds to accidental HOM ignition, while the next peak corresponds to the ignition of the plasma in the cavity, followed by a rapid sequence of smaller intensity peaks that identify the plasma being moved from cell $\#$ 5 to $\#$ 9 and then tuned in cell $\#$ 9. After approximately one hour, the plasma density is decreased, and the glow discharge is moved to the next cell. This procedure is repeated for each cell. The temperature increase measured on the HOM1 and HOM2 connectors of all four plasma-processed cavities is shown in panel (c). In the period plotted here, the glow discharge was ignited in CAV4, as confirmed by the temperature increase registered by its HOM1 and HOM2 sensors. Panel (d) shows the temperature increase measured on the outside of cells $\#$ 1 and $\#$ 9. The four sensors are placed inside the He vessel at the top and bottom of the end-cells.}
	\label{fig:CAV4plots}
\end{figure*}

Figures \ref{fig:GasPressure_cav4} and \ref{fig:CAV4plots} show a complete set of data collected during the first plasma cleaning run on CAV4. This cavity is used as an example to explain the data monitored during plasma processing and their typical trends and values. 
Figure \ref{fig:GasPressure_cav4} shows the pressure data acquired by the gauges at the two ends of the cryomodule at the gas injection and vacuum sides. This graph shows the pressure increase at both
ends as the flow of gas is initiated into the beamline, the pressure stabilization phase and the pressure drop due to the beamline evacuation at the end of the processing. The pressure difference between the gas injection side and the vacuum side indicates the pressure drop between the two ends.

Figure \ref{fig:CAV4plots} shows different signals recorded during plasma processing, and it also illustrates a rare case in which plasma ignition occurred at the HOM. Figure \ref{fig:CAV4plots}(a) shows the partial pressure as a function of time of some of the RGA signals measured on CAV4. Only the signals relevant for plasma processing are plotted: oxygen, neon, carbon-related peaks, and water. The initial peak in oxygen is always observed by the RGA as soon as the gas flow is established in the cavity string. This peak quickly decreases, and the oxygen concentration reaches equilibrium. The first peak in \ch{CO}, \ch{CO2} is due to the plasma being accidentally ignited at the HOM coupler rather than inside the cavity volume. This event, which we refer to as HOM ignition, lasted only a few seconds (\SI{5}{\second} according to the power levels shown in Fig. \ref{fig:CAV4plots}(b)). As reported in the next section, the results of the RF test after plasma cleaning confirmed that this event exerted no negative effect on the cavity performance. However, it is best practice to avoid HOM ignition, as prolonged plasma ignition in the coupler could cause the formation of an arc between the antenna and the cavity, and could potentially cause sputtering of particles into the cavity. For this reason, after the first HOM ignition, we collected additional $S_{21}$, $S_{11}$ and $S_{22}$ measurements on CAV4 with the VNA to optimize the plasma ignition technique for this particular cavity. The additional measurements provided a better understanding of the coupling of the HOM1 and HOM2 antennas to the 2D-1 mode (second dipole pass-band, first resonant mode) used for plasma ignition and allowed us to successfully ignite the plasma inside the cavity volume, as shown by the small, broad peak in \ch{CO}, \ch{CO2} in the RGA plot (approximately 11:00 in Fig. \ref{fig:CAV4plots}(a)). No other cell in this cavity showed peaks in the carbon-related signals when the plasma was ignited.

The input, reflected and transmitted power (respectively $\mathrm{P_{IN}}$, $\mathrm{P_{RFL}}$, $\mathrm{P_{T}}$) measured during the first run of plasma processing on CAV4 are shown in Fig. \ref{fig:CAV4plots}(b). The first peak at approximately 9:30 corresponds to HOM ignition, which was reached at $\mathrm{P_{IN}}=\SI{94.2}{\watt}$, $\mathrm{P_{RFL}}=\SI{19.1}{\watt}$, and $\mathrm{P_{T}}=\SI{1.3}{\milli\watt}$ and lasted $\approx \SI{5}{\second}$. Just before 11:00, the plasma was ignited inside the cavity volume in cell $\#$ 5, requiring $\mathrm{P_{IN}}=\SI{67.3}{\watt}$, $\mathrm{P_{RFL}}=\SI{14.9}{\watt}$, and $\mathrm{P_{T}}=\SI{1.4}{\milli\watt}$. Once ignited using mode 2D-1, mode 1D-5 (first dipole pass-band, fifth resonant mode) is quickly added using the second RF signal generator, and the first high power mode is turned off. At this point, the plasma is maintained using only $\mathrm{P_{IN}}=\SI{6.5}{\watt}$. The following cluster of peaks at approximately 11:00 corresponds to the plasma being transferred from cell $\#$ 5 to cell $\#$ 9 and tuned to the desired plasma density. After almost an hour, just before 12:00, the plasma is detuned, meaning that its density is reduced, and it is moved to the following cell, where its density is once again maximized. This process is repeated for all the cells. Every time the plasma is moved from one cell to the other, the detection program is used to confirm the cell of ignition. Each cell is processed for approximately \SI{1}{\hour} and typically $\mathrm{P_{IN}} \approx \SI{10}{\watt}$, $\mathrm{P_{RFL}} \approx \SI{5}{\watt}$, $\mathrm{P_{T}} \approx \SI{0.15}{\milli\watt}$. When a combination of two modes is used to transfer the plasma or to tune its density, the input power can temporarily reach $\mathrm{P_{IN}} \approx \SI{20}{\watt}$.

Figure \ref{fig:CAV4plots} panels (c) and (d) show the temperature increase measured for the HOM cable connectors and on the cavity external surface (on cell $\#$ 1 and 9) during CAV4 plasma processing. In Fig. \ref{fig:CAV4plots}(c), the CAV4 HOM1 sensor shows an increase in temperature corresponding first to the accidental ignition of the plasma in the HOM ($\Delta T = \SI{1.4}{\kelvin}$) and later to the ignition of the plasma in the cavity. A new spike in temperature is registered when two modes are superimposed to increase (or decrease) the plasma density at the beginning (or end) of each cell processing and to move it to a new cell. As expected, the HOM2 cable connector exhibits a moderate temperature increase with no particular trend connected to the position of the plasma in the cavity. A plot of the temperature measured on the CAV4 external surface of cells $\#$ 1 and 9 is shown in Fig. \ref{fig:CAV4plots}(d). The four sensors are placed inside the helium vessel and at the top and bottom of cells $\#$ 1 and 9. When the plasma is ignited in cell $\#$ 9, the temperature sensors on this cell measure a maximum $\Delta T \approx \SI{1}{\kelvin}$. As the plasma is moved farther away from cell $\#$ 9, the $\Delta T$ measured on these sensors decreases, while the temperature measured on cell $\#$ 1 gradually increases up to $\Delta T \approx \SI{1.5}{\kelvin}$. Although clearly dependent on plasma processing, cable and cavity heating are not a source of concern, since the total cable heating was $\Delta T_{HOM1} \approx \SI{2}{\kelvin}$, $\Delta T_{HOM2} \approx \SI{0.5}{\kelvin}$ and the maximum temperature increase on the cavity surface was $\Delta T \approx \SI{1.5}{\kelvin}$.

\begin{figure*}
    \centering
    \includegraphics[width=0.9\textwidth]{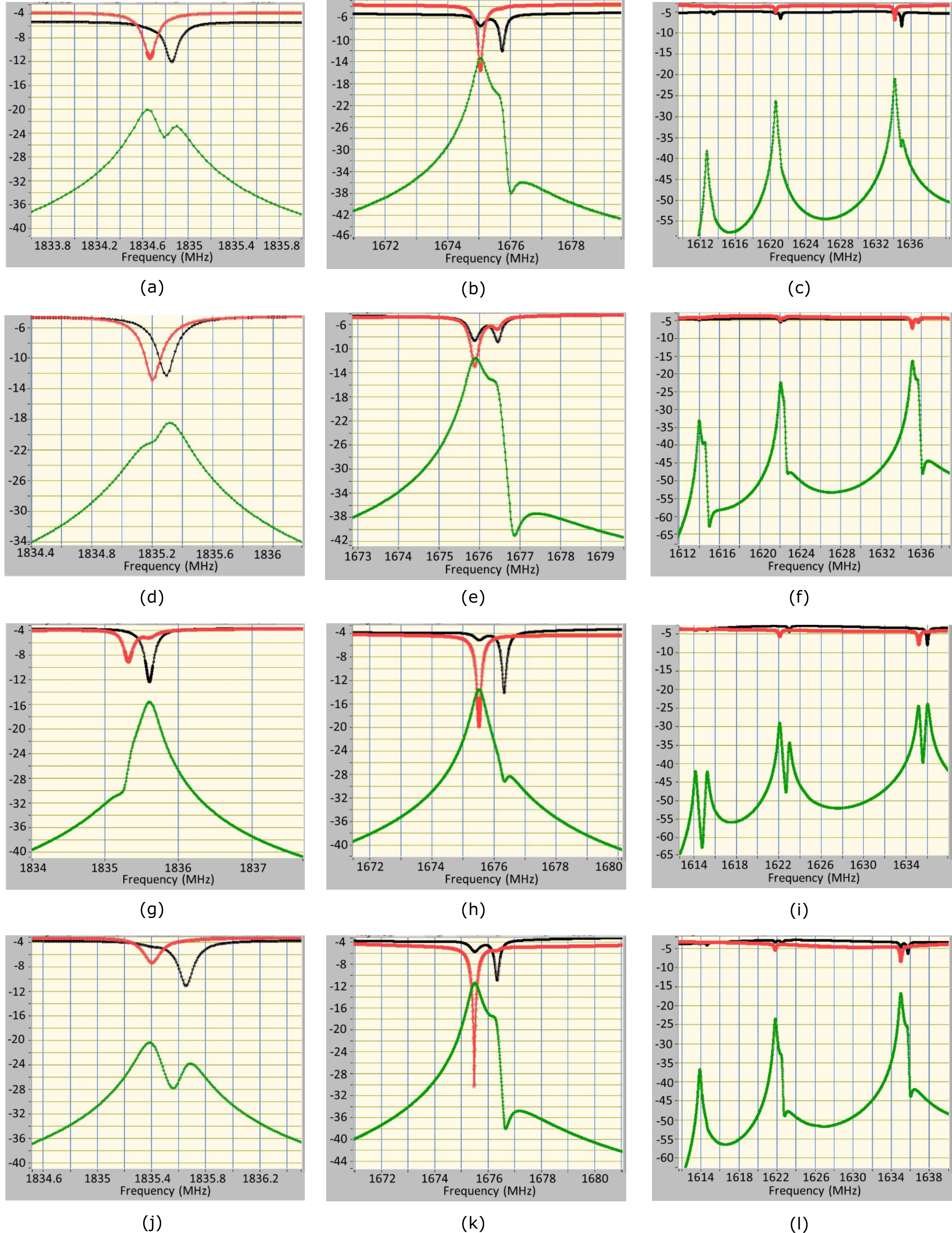}
    \caption{Plots of the $S_{21}$ (green), $S_{11}$ (red), and $S_{22}$ (black) parameters for some of the dipole modes used for plasma processing. The parameters were measured on CAV1-4-5-8 from top to bottom using a network analyzer. Panels (a), (d), (g), and (j) contain the 2D-1 peaks, panels (b), (e), (h), and (k) show the 1D-5 peaks, and panels (c), (f), (i), and (l) show the 1D-1, 1D-2, and 1D-3 peaks. The plots present an example of the different couplings to the HOM and the dipole splitting characteristics of each cavity.}
    \label{fig:S21par}
\end{figure*}

The plots discussed in Fig. \ref{fig:CAV4plots} all report data collected during the first run of plasma processing applied to CAV4. All four plasma-processed cavities showed an initial peak in carbon-related signals on the RGA when the plasma was ignited for the first time inside the cavity and moved to cell $\#$ 9. Only CAV1 showed additional peaks in cell $\#$ 1, 4 and 5 and a small peak in cell $\#$ 9 during the second pass of plasma cleaning. In all cases, the carbon-related peaks quickly decreased back to the background level. The plasma was accidentally ignited in the HOMs of CAV4 and CAV8. In both cases, the HOM ignition lasted only a few seconds and only occurred once. In both cavities, during the second attempt, the glow discharge was successfully ignited inside the cavity RF volume by carefully tuning the RF driving frequency to match the HOM1 optimal coupling to the 2D-1 mode. The plasma processing method for LCLS-II cavities uses higher order modes, which confers the advantage of offering good coupling even at room temperature, making it possible to ignite the plasma with a few ten of watts and to then maintain it with only $5-\SI{10}{\watt}$. On the other hand, contrary to the fundamental pass-band, the HOMs are not tuned. The HOM couplers are only tuned so that their notch filters reject the fundamental mode. This lack of tuning of higher order modes implies that the frequencies of each higher order mode, the splittings in dipole modes and the couplings to the HOM antennas vary from cavity to cavity. In the four cavities processed in the vCM, different configurations of degenerate dipole splittings were found, as shown in Fig. \ref{fig:S21par}. Measurements of the $S_{11}$ and $S_{22}$ parameters for the HOMs allowed us to identify the exact frequency to send in input to each cavity to successfully ignite the plasma in the central cell using mode 2D-1. This, along with the work conducted during the R$\&$D phase, demonstrated that the ignition of the plasma through the HOM is a robust method that guarantees successful ignition and control of the glow discharge inside the cavity RF volume.

Once the processing of the four cavities was completed, the gas cart was isolated from the string, which was evacuated up to $\approx 1 \times 10^{-6}\,$Torr using the plasma vacuum cart, and then the CMTF ion pumps were turned on. Both gas and vacuum carts were later disconnected from the vCM using a portable class-100 cleanroom.

\section{\label{sec:vCMResults} Comparison of the vCM performance before and after plasma processing}

After plasma processing and evacuation, the vCM was cooled down to be RF tested again. The purpose of this test was to verify that plasma processing did not cause any deterioration in cavity performance and to understand if it had an impact on MP. The plan was to remeasure the quality factor, the maximum and the usable gradient for each cavity, the presence of X-rays and dark current and the stability of operations at $20.8\,$MV/m. The time necessary to process multipacting was also investigated.

The results of the vCM prior to plasma processing are reported in table \ref{tab:vCMPerformance}, along with the comparison with the performance measured after plasma cleaning. For all eight cavities, it was verified that the performance was preserved both in terms of quality factor and accelerating gradient. This demonstrates that plasma processing did not introduce any observably detrimental contamination or particulates inside the cavity string, maintaining the vCM field emission-free. After plasma processing, the CM test confirmed the total module voltage of \SI{210}{\mega\volt} versus the \SI{173}{\mega\volt} required by the LCLS-II-HE specification. The vCM average quality factor still exceeded the specification ($Q_0 = 2.7 \times 10^{10}$ at $20.8\,$MV/m), with an average of $Q_0 = 3.1 \times 10^{10}$. As explained in Posen \textit{et al.} \cite{posen2021lcls}, the administrative limit for the gradient in case the ultimate quench field of the cavity was not reached was set to $26.0\,$MV/m, while the maximum gradient defines the field level at which a cavity would quench consistently without allowing any further field increase. The usable gradient instead is defined as the maximum field at which i) the cavity can operate for more than one hour without quenching, ii) the radiation remains below $50\,$mR/hr and iii) the cavity is 0.5 MV/m below its ultimate quench field limit. 

\begin{table*}
    \caption{\label{tab:vCMPerformance}Comparison of vCM performance measured before and after plasma processing. The four cavities to which plasma processing was applied are highlighted in bold.}
\begin{ruledtabular}
\begin{tabular}{ccccccccc}
& \multicolumn{4}{c}{Before plasma} & \multicolumn{4}{c}{After plasma} \\
Cavity & Max $\mathrm{E_{acc}}$ & Usable $\mathrm{E_{acc}}$ & $\mathrm{Q_0}$ at $21\,$MV/m & MP quenches & Max $\mathrm{E_{acc}}$ & Usable $\mathrm{E_{acc}}$ & $\mathrm{Q_0}$ at $21\,$MV/m & MP quenches \\
& (MV/m) & (MV/m) & $\times 10^{10}$ & & (MV/m) & (MV/m) & $\times 10^{10}$ & \\ \hline
\textbf{1} & \textbf{23.4} & \textbf{22.9} & \textbf{3.0} & \textbf{Yes} & \textbf{23.8} & \textbf{23.3} & \textbf{3.4} & \textbf{No} \\
2 & 24.8 & 24.3 & 3.0 & Yes & 25.2 & 24.7 & 3.2 & Yes \\
3 & 25.4 & 24.9 & 2.6 & Yes & 26.0 & 26.0 & 3.4 & Yes \\
\textbf{4} & \textbf{26.0} & \textbf{26.0} & \textbf{3.2} & \textbf{Yes} & \textbf{26.0} & \textbf{26.0} & \textbf{3.2} & \textbf{No} \\
\textbf{5} & \textbf{25.3} & \textbf{24.8} & \textbf{2.9} & \textbf{Yes} & \textbf{25.5} & \textbf{25.0} & \textbf{2.8} & \textbf{No} \\
6 & 26.0 & 25.5 & 3.4 & Yes & 26.0 & 26.0 & 3.2 & Yes \\
7 & 25.7 & 25.2 & 3.4 & Yes & 25.9 & 25.4 & 3.3 & Yes \\
\textbf{8} & \textbf{24.4} & \textbf{23.9} & \textbf{2.7} & \textbf{Yes} & \textbf{24.7} & \textbf{24.2} & \textbf{2.6} & \textbf{No} \\ \hline
Average & 25.1 & 24.7 & 3.0 & & 25.3 & 25.1 & 3.1 & \\
Total & 209 & 205 & & & 210 & 208 & & \\
\end{tabular}
\end{ruledtabular}
\end{table*}

As shown in Fig.$\,$(10) and (11) from Posen \textit{et al.} \cite{posen2021lcls}, during the first vCM RF test, all eight cavities suffered from multipacting induced quenches during both the power rise to $21\,$MV/m (Fig. (10) \cite{posen2021lcls}) or the long duration operation at high gradient (Fig. (11) \cite{posen2021lcls}). As highlighted by the last column in table \ref{tab:vCMPerformance}, after plasma cleaning the four cavities that were not processed still showed MP quenches, while the four processed cavities did not experience any MP-related quench. 

In table \ref{tab:vCM_MP}, the numbers of quenches attributed to MP for each cavity, as measured during three different cooldowns, are reported. The first two cooldowns were both measured during the initial test of the vCM. No data of MP quenches for CAV1 relative to the first cooldown are available, as initially there was a loose connection in the coupler that was later secured when the cryomodule was warmed to room temperature. The After Plasma column shows the multipacting quenches after plasma cleaning of the vCM cavities. For all three RF tests, the data includes the MP quenches encountered during the initial power rise to $16\,$MV/m, the power rise to the maximum accelerating gradient (often accompanied by MP processing as necessary), the measurement of the usable gradient followed by additional MP processing, the stability test at $21\,$MV/m (usually \SI{1}{\hour} or more), and additional MP processing as necessary. The MP quenches that took place during pulsed RF processing (applied to CAV3 and CAV5 during the first cooldown and to CAV1 and CAV6 during the second cooldown) are not counted by our algorithm, causing underestimation in the number of quenches estimated for the two pre-plasma cooldowns. 

\begin{figure*}
    \centering
    \includegraphics[width=1\textwidth]{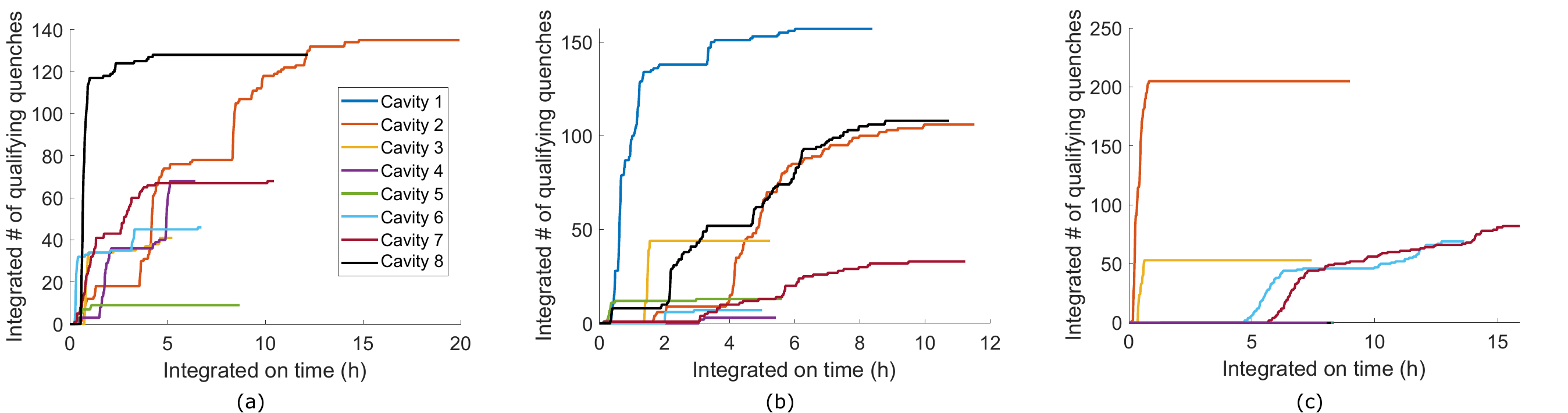}
    \caption{The three plots show the number of MP quenches found for each cavity as a function of the integrated time that the cavity spent powered on. Plot (a) shows these results for the $1^{st}$ cooldown, (b) for the $2^{nd}$ cooldown measured before plasma processing and (c) for the after plasma cooldown. }
    \label{fig:quenchesnumber}
\end{figure*}

 To identify and quantify the MP quenches the data concerning the evolution of the cavity gradient and radiation level as a function of time were analyzed for each cavity. An algorithm was used to identify all the quenches, seen as a sudden drop in the gradient signal, above $16\,$MV/m and below the maximum accelerating field, such as due to MP whenever a radiation signal above background was detected within the two minutes around the quench event. For this purpose, all the radiation sensors were considered, not only the sensors closest to the cavity under examination. 
 
 As explained in Posen \textit{et al.} \cite{posen2021lcls}, CAV5 was affected by FE during the first two cooldowns here under consideration. The electron activity caused by the field emission may have contributed to CAV5 MP-induced quenches during the first two cooldowns. At the end of the unit test, the FE source was processed, leaving the CAV5 (as the rest of the vCM) field emission-free. After the unit test, CAV5 still showed a few multipacting quenches. A higher radiation threshold was used in the cases when CAV5 was on and exhibiting FE while other cavities were powered on at the same time.
 
 One limitation of this methodology, in general, is due to the fact that in case multiple quenches occur close in time, it is not possible to distinguish the radiation produced by each quench event.
 
 Figure \ref{fig:quenchesnumber} shows the number of MP quenches measured for each cavity as a function of the integrated time during which the cavity was tested for the three cooldowns under study. Additionally, from the graph in panel \ref{fig:quenchesnumber}(c), it is possible to notice that the four plasma-cleaned cavities did not experience any MP-related quench during the RF test, while the four non-processed cavities still showed MP quenches.

\begin{table}
    \caption{\label{tab:vCM_MP}Number of multipacting quenches measured for each cavity in two different cooldowns prior to plasma processing and the one cooldown after plasma cleaning. The four cavities to which plasma processing was applied are highlighted in bold.}
\begin{ruledtabular}
\begin{tabular}{cccc}
Cavity & \multicolumn{3}{c}{Multipacting Quenches} \\
 & \multicolumn{2}{c}{Before plasma} & After Plasma\\
 & $\mathrm{1^{st}}$ cooldown &  $\mathrm{2^{nd}}$ cooldown & \\  \hline

\textbf{1} & / & \textbf{157} & \textbf{0} \\
2 & 135 & 106 & 205 \\
3 & 41 & 44 & 53 \\
\textbf{4} & \textbf{68} & \textbf{3} & \textbf{0} \\
\textbf{5} & \textbf{10} & \textbf{16} & \textbf{0} \\
6 & 46 & 7 & 69 \\
7 & 68 & 33 & 82 \\
\textbf{8} & \textbf{128} & \textbf{108} & \textbf{0} \\
\end{tabular}
\end{ruledtabular}
\end{table}

This encouraging result demonstrated that although the plasma cleaning methodology was developed and optimized to target field emission, it is also effective in eliminating multipacting of cavities in cryomodules. Indeed, it is known that both field emission and multipacting can be enhanced by the presence of adsorbates on the cavity surface \cite{padamseerf}. For example, hydrocarbons can reduce the Nb work functions, decreasing the FE threshold, while the presence of water can increase the Nb secondary emission yield, increasing MP. Plasma processing can successfully decrease the level of different types of adsorbates, mitigating both FE and MP \textit{in situ} at the same time. Previous studies conducted at FNAL during the R$\&$D phase used visible carbon-based contamination to study the reach of the plasma inside the cavity cell and showed that the removal is maximal at the iris but still present and nonnegligible at the equator. More fundamental studies will follow to better understand how to optimize plasma processing for multipacting mitigation.

\section{\label{sec:conclusions} Conclusions}

A plasma cleaning methodology was fully developed for the Linac Coherent Light Source II (LCLS-II) cavities. We presented its first application to a full cryomodule system, i.e., to the LCLS-II High Energy (HE) verification cryomodule.

During the developmental phase, we reviewed and mitigated possible risks deriving from the application of plasma processing to the cryomodule. This phase was crucial to confirm that the procedure could be safely applied to the cryomodule without the risk of compromising its components. 

After connecting all the hardware needed for plasma cleaning, four out of eight cavities of the vCM were fully processed. The cryomodule was subsequently cooled down and retested to verify the effect of the processing on the cavity performance.
With this test, we demonstrated that after plasma cleaning, i) CM integrity and cavity performance are preserved, ii) no field emitters were introduced by the processing (the vCM is still field emission-free), and iii) plasma processing eliminates multipacting: plasma processed cavities did not show any sign of MP, while the other four cavities were again affected by a series of MP quenches, also observed during previous testing.
Through this test, the plasma cleaning procedure was fully validated, showing that plasma processing has the potential to not only reduce field emission but also fully eliminate the MP of cavities in CMs. 

More fundamental studies will be needed to better understand how to optimize the plasma processing procedure for in situ multipacting mitigation. In addition, the plasma cleaning procedure may be further optimized to be easily implemented during the LCLS-II-HE cryomodule production phase. This could significantly reduce cryomodule testing and commissioning time and increase reliability during machine operation.

\begin{acknowledgments}
The authors would like to express their gratitude to Dr. Eliane Lessner and the Office of Basic Energy Sciences for supporting the development of plasma processing for LCLS-II and LCLS-II-HE cryomodules.

The authors would like to thank LCLS-II-HE for funding the application of plasma processing to the \SI{1.3}{\giga\hertz} 9-cell LCLS-II-HE verification cryomodule.

The authors would like to also thank Elias Lopez for assisting and supporting the plasma processing operations at the FNAL Cryomodule Test Facility. 

This work was supported by the United States Department of Energy (DOE), Offices of High Energy Physics and Basic Energy Sciences under Contracts DE-AC05-06OR23177 (Fermilab), DE-AC02-76F00515 (SLAC), and DE-AC05-00OR22725 (ORNL).
\end{acknowledgments}

\bibliography{vCM_PlasmaPaper}

\end{document}